\documentclass[aps,pra,reprint,twocolumn,groupedaddress,longbibliography,floats,final]{revtex4-2}
\usepackage[usenames,dvipsnames]{xcolor}
\usepackage{graphicx}
\usepackage{amssymb}
\usepackage{subfigure}
\usepackage{stackrel}
\usepackage{amsmath}
\usepackage{amsthm}
\usepackage{amsfonts}
\usepackage{bm}
\usepackage{graphicx}
\usepackage{mwe}
\usepackage[normalem]{ulem}
\usepackage{longtable}
\usepackage{float}
\usepackage{comment}

\newtheorem*{theorem*}{Assertion}
\renewcommand{\qedsymbol}{$\blacksquare$}

\definecolor{rosita}{rgb}{0.97, 0.56, 0.65}
\usepackage{amsmath}
\newcommand{\beq}{\begin{equation}}
\newcommand{\eeq}{\end{equation}}
\newcommand{\ket}[1]{\ensuremath{\left|{#1}\right\rangle}}

\newcommand{\braket}[2]{\ensuremath{\langle{#1}|{#2}\rangle}}
\newcommand{\op}[1]{\ensuremath{\hat{\mathnormal{#1}}}}
\newcommand{\mean}[1]{\ensuremath{\langle{#1}\rangle}}

\begin{document}

\title{Speed of evolution in qutrit systems}

\author{Jesica \surname{Espino-Gonz\'alez}}
\affiliation{Instituto de F\'{\i}sica, Universidad Nacional Aut\'{o}noma de M\'{e}xico,
Apartado Postal 20-364, Ciudad de M\'{e}xico, Mexico.}

\author{Francisco J. \surname{Sevilla}}
\affiliation{Instituto de F\'{\i}sica, Universidad Nacional Aut\'{o}noma de M\'{e}xico, Apartado Postal 20-364, Ciudad de M\'{e}xico, Mexico.}
\email[]{fjsevilla@fisica.unam.mx}
\thanks{}

\author{Andrea Vald\'es-Hern\'andez}\email{Corresponding author: andreavh@fisica.unam.mx}
\affiliation{Instituto de F\'{\i}sica, Universidad Nacional Aut\'{o}noma de M\'{e}xico,
Apartado Postal 20-364, Ciudad de M\'{e}xico, Mexico.}

\begin{abstract}

The speed of evolution between perfectly distinguishable states is thoroughly analyzed in a closed three-level (qutrit) quantum system. 
Considering an evolution under an arbitrary time-independent Hamiltonian, we fully characterize the relevant parameters according to whether the corresponding quantum speed limit is given by the Mandelstam-Tamm, the Margolus-Levitin, or the Ness-Alberti-Sagi (dual) bound, 
thereby elucidating their hierarchy and relative importance. 
We revisit the necessary and sufficient conditions that guarantee the evolution of the initial state towards an orthogonal one in a finite time, and pay special attention to the full characterization of the speed of evolution, offering a speed map in parameter space that highlights regions associated with faster or slower dynamics.
The general analysis is applied to concrete physical settings, particularly a pair of bosons governed by an extended Bose-Hubbard Hamiltonian, and a single particle in a triple-well potential. Our findings 
provide a framework to explore how the energetic resources and the initial configurations shape the pace of the dynamics in higher-dimensional systems.

\end{abstract}
\maketitle

\section{Introduction}
The fundamental limits on the time required for a quantum state to evolve to an orthogonal one have been a subject of intense study, leading to a whole research field 
 known as \emph{Quantum Speed Limit} (QSL)\cite{DeffnerJPA2017,FreyQIP2016}. 
In its original context, the QSL determines the minimal time, imposed by the system's energetic resources, to continuously transit between orthogonal states. 
It establishes a naturally intrinsic time-scale of the dynamics,  characterizes the evolution rate of the system, and stands as a central quantity for many processes in quantum computation.
%

For pure states $\ket{\psi}$ evolving under a time-independent Hamiltonian $\op{H}$, three lower bounds have been established that limit the orthogonality time $\tau$, required to actually reach an orthogonal state.
The first bound is based on a time-energy uncertainty relation,
rigorously formulated by Mandelstam and Tamm (MT)\cite{Mandelstam1991},
who showed that $\tau$ is 
bounded from below by the inverse
of the system's energy dispersion $\sigma_{\op{H}}$, giving rise to the MT bound, $\tau_{\textrm{MT}}=\pi\hbar/(2\sigma_{\op{H}})$.
The second bound was put forward by Margolus and Levitin (ML) \cite{MargolusPhysicaD1998} and is given by $\tau_{\textrm{ML}}=\pi\hbar/(2\mathcal E)$. 
It stands out by exhibiting that the mean energy $\mathcal E$, referred to the minimal energy $E_{\min}$ that contributes to $\ket{\psi}$ (when expressed in the energy representation), determines a lower bound to $\tau$.
More recently, Ness, Alberti and Sagi found a bound, dual to the ML one and given by $\tau_{\textrm{ML*}}=\pi\hbar/(2\mathcal E^*)$ \cite{NessPhysRevLett2022}, which restricts the orthogonality time in terms of $\mathcal E^*$, the mean energy referred to the energy of the highest occupied eigenstate. 

The intrinsic limitations on the temporal evolution of quantum systems have been approached from different fronts and extended to different trends \cite{FreyQIP2016,DeffnerJPA2017,PiresPhysRevX2016,OkuyamaPhysRevLett2018,MohamPhysRevA2022,WangCommunTheorPhys2022,ZhangNPJQuantInf2023,WuPhysRevA2023,JohnssonPhysRevA2023,carabbaBook2024,ChauJPhysA2024}.  
Distinct theoretical frameworks have been introduced for the analysis of the QSL in pure states considering the evolution towards orthogonal states \cite{AnandanPRL1990,VaidmanAJP1992,SoderholmPhysRevA1999}, or towards pure states with a given finite fidelity \cite{Giovannetti2003PRA,HegdePhysRevLett2022,HornedalPhysRevResearch2023,AndrzejewskiQIP2024,Osan2025}. 
Important advances on the study of the QSL for mixed states have been obtained \cite{DeffnerJPhysA2013,ZhangSciReps2014,MondalPhysLett2016a,MondalPhysLett2016b,CampaioliPhysRevLett2018,HornedalNewJPhys2022,BagchiEntropy2023}, and similarly for open quantum systems \cite{TaddeiPRL2013,DeffnerPRL2013,delCampoPRL2013,SunSciReps2015,CampaioliQuantum2019}.  
Further, different quantum speed limits have been introduced for non-Hermitian \cite{BenderPRL2007,SunPhysRevLett2021} and perturbed \cite{YadinPhysRevLett2024} systems.
 Other generalizations of speed limits consider the speed of evolution of transformation 
 implementations \cite{PoggiPhysRevA2019,FarmanianPhysRevA2024}, and of  
 different quantum resources \cite{MarvianPhysRevA2016,CampaioliNewJPhys2022,MohanNewJPhys2022,ChenRepProgPhys2023,HerbPhysRevLett2024,Naseri_NJP_2024}.

The experimental verification of the QSL has been challenging and recent progress has been made in this direction. A seminal experimental proposal to measure the QSL in a system of ultracold atoms confined in a time-dependent harmonic trap 
was presented in Ref.~\cite{delCampoPRL2021} while more recent proposals have realized the QSL in different systems \cite{PiresCommPhysics2024,YaouzuPhysRevA2024,ZhuNatCoomun2025}.

Despite all the advances achieved in the study of quantum speed limits, the identification of the set of states that actually attain an orthogonal state, along with the states that do it faster, stands as an open and challenging problem that deserves attention \cite{zhangNPJQInf2023}. 
Indeed, such analysis acquires relevance in determining the specific type of evolution (optical control theory \cite{CanevaPhysRevLett2009}) and preparation of the states that favor a faster evolution into an orthogonal state.   
For isolated two-level (qubit) systems, a distinguishable state is attained irrespectively of the time-independent evolution Hamiltonian, provided the initial state decomposes as an equally weighted superposition of two (non-degenerate) energy eigenstates \cite{LevitinPRL2009}. 
In this case the evolution \emph{is the fastest one} allowed by the QSL.  
Three-level (qutrit) systems are therefore the simplest ones in which the conditions that guarantee the transit between distinguishable states becomes non-trivial. Qutrit systems constitute a relevant platform for many quantum tasks as in: quantum computation with trapped ions \cite{KlimovPhysRevA2003,McHughNJPhys2005}, quantum information transfer \cite{DelgadoPhysLettA2007}, the implementations of specific quantum channels \cite{VitanovPhysRevA2012,YurlatanPhysRevLett2020}, and have also been engineered in quantum optical systems \cite{BogdanovPhysRevLett2004} and in ultracold atom systems \cite{LindonPhysRevApplied2023}. 

A first step for studying the speed of evolution towards a distinguishable state in qutrit systems   was initiated in \cite{SevillaQRep2021,KhanQInfProc2021}. In Ref.~\cite{SevillaQRep2021}
the orthogonality condition was exactly solved considering a (time-independent) Hamiltonian evolution, and a complete characterization of the relevant parameters was advanced. 
Yet, the analysis involved only the MT and the ML bounds (leaving out the recently found dual bound ML*), and left open the problem of determining the actual speed of evolution between distinguishable states.
In the present work we consider all three bounds on equal footing, with the double purpose of: i) fully characterize the QSL by identifying the regions (in the parameters space) in which either bound becomes dominant, and ii)  
thoroughly analyse the speed of evolution towards an orthogonal state, enabling the identification of the conditions, both on the preparation state and the Hamiltonian spectrum, that distinguish slower from faster states. 
%

In Section \ref{sect:QSL} we revisit the notion of quantum speed limit as the unified bound involving the MT, the ML, and the dual bound, and introduce suitable quantities that define the QSL in a qutrit system.
Then, in Section \ref{mapaQSL}, regions in the parameter space are identified according to the specific bound that determines the QSL, and a map of the QSL in qutrit systems is depicted.  
In Section \ref{sect:Orthogonality} we revisit the set of initial states that reach a distinguishable one in a finite time and compute the actual orthogonality time. 
By comparison of the latter with the corresponding QSL we 
analyse the speed of evolution in Section \ref{sect:MapSpeeds}, 
and construct a comprehensive map of evolution speed across the parameter space, thereby identifying the initial state configurations and the system energetic resources that lead to faster, or slower, evolution.
In Section \ref{physical} we apply the results of our analysis to concrete physical systems, particularly to a pair of bosons considering an extended Bose-Hubbard model.


\section{\label{sect:QSL} Quantum Speed Limit}


By consideration of the MT, ML and ML$^*$ bounds, the quantum speed limit for a pure state $\ket{\psi}$ evolving under a time-independent Hamiltonian $\hat H$ is given by the unified generalized bound \cite{NessPhysRevLett2022}
\begin{eqnarray}
\label{BoundN}
\tau_{\text{\textrm{qsl}}}&\equiv& \max\left\{\tau_{\textrm{MT}},\tau_{\textrm{ML}},\tau_{\textrm{ML*}}\right\}\nonumber\\
&=&\max\left\{\frac{\pi\hbar}{2\sigma_{\op{H}}},\frac{\pi\hbar}{2\mathcal{E}},\frac{\pi\hbar}{2\mathcal {E}^{*}}\right\},
\end{eqnarray}
where $\sigma_{\op{H}}=\sqrt{\mean{\op{H}^{2}}-\mean{\op{H}}^{2}}$, $\mathcal{E}=\mean{\op{H}}-E_{\min}$, and $\mathcal{E}^*=E_{\max}-\mean{\op{H}}$, with $E_{\min(\max)}$ the minimal(maximal) energy that contributes to $\ket{\psi}$, when the state is expanded in the Hamiltonian eigenbasis.

Clearly the QSL is completely determined by  
the ratios 
\beq\label{cocientes}
\alpha\equiv\frac{\tau_{\textrm{ML}}}{\tau_{\textrm{MT}}}=\frac{\sigma_{\op{H}}}{\mathcal{E}},\quad \beta\equiv\frac{\tau_{\textrm{ML*}}}{\tau_{\textrm{MT}}}=\frac{\sigma_{\op{H}}}{\mathcal {E}^{*}},
\eeq
in such a way that
\beq\label{cotas}
\tau_{\text{qsl}}=\begin{cases}
			\tau_{\textrm{MT}}  \iff \alpha<1,\beta<1,\\
            \tau_{\textrm{ML}} \iff \beta<\alpha,1<\alpha,\\
           \tau_{\textrm{ML*}} \iff \alpha<\beta,1<\beta.
		 \end{cases}
   \eeq
%

\begin{figure}
\includegraphics[width=0.75\columnwidth]{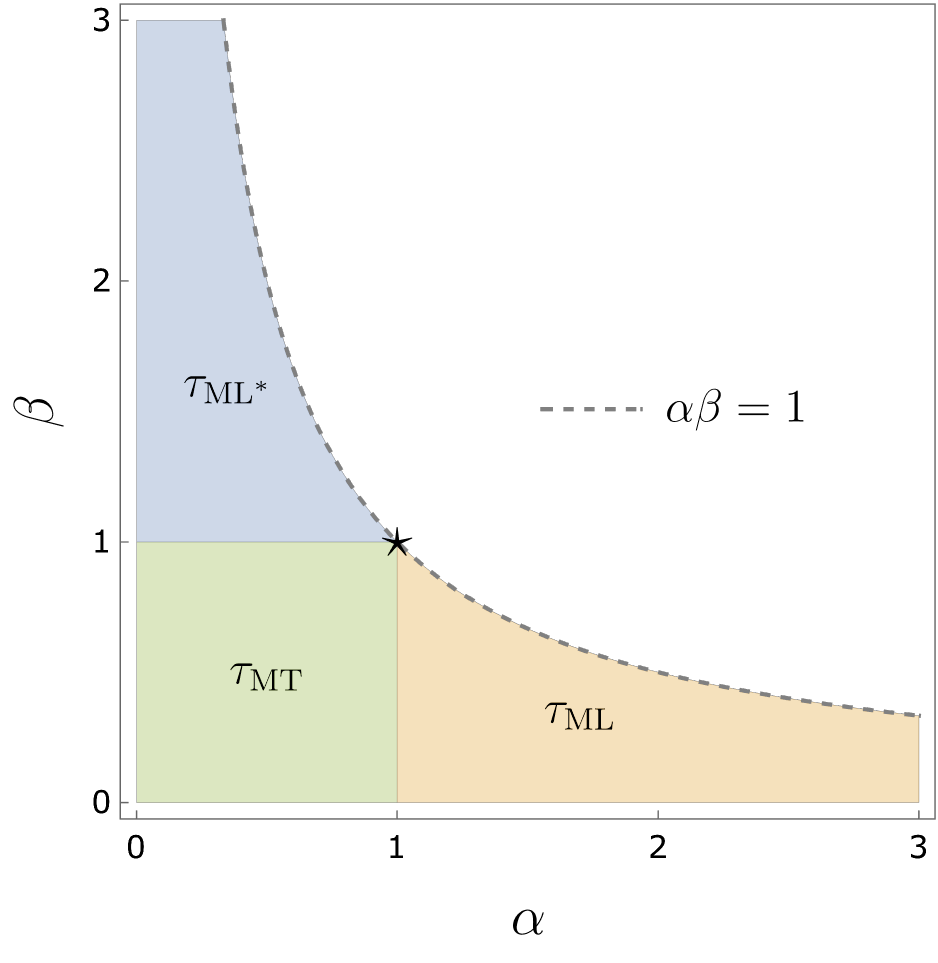}
\caption{Region in the space $(\alpha,\beta)$ that is consistent with the Bhatia-Davis inequality decomposes into three (colored) regions, each one delimited according to (\ref{cotas}), thus determining the QSL as either one of the bounds $\tau_{\textrm{MT}}$, $\tau_{\textrm{ML}}$, and $\tau_{\textrm{MT*}}$.
}
\label{fig:alpha-beta}
\end{figure}

The Bhatia-Davis inequality \cite{BhatiaAmMathMonthly2000}, namely $\sigma_{\op{H}}^2\le\mathcal{E}\mathcal{E}^*$, 
can be written straightforwardly in terms of $\alpha$ and $\beta$ as 
$0\le\alpha\beta\le1$. 
The corresponding region in the space $(\alpha,\beta)$ is depicted in Fig.~\ref{fig:alpha-beta}, with the colors indicating the zones in which the QSL is given by either one of the bounds MT, ML or ML*.
The star highlights the single point ($\alpha=\beta=1$) at which the three bounds coincide, and corresponds to states $\ket{\psi}$ that decompose as an equally-weighted superposition of only two non-degenerate energy eigenstates \cite{LevitinPRL2009}. 
The dashed line ($\alpha\beta=1$) contains the unbalanced superpositions of two non-degenerate energy eigenstates, whence effective qubit states lie along the boundary of the colored regions in Fig. \ref{fig:alpha-beta}.
For such systems, only two bounds become relevant, ML and ML*. 
Therefore, as has been previously pointed out in \cite{NessPhysRevLett2022}, a higher dimensional Hilbert space is required for the three bounds to become non-trivially determined. 

In the following sections we will thoroughly explore the map of the QSL in the relevant parameters space, considering the simplest non-trivial scenario, namely a three-level (qutrit) system, which has been realized experimentally in optical and ultracold atoms systems \cite{LindonPhysRevApplied2023}. 

\subsection{$\alpha$ and $\beta$ in a qutrit system}\label{Aalphabeta}
The initial qutrit pure state under consideration decomposes as
\begin{equation}\label{Psi0}
\ket{\psi_0}=\sum_{i=1}^{3}\sqrt{r_{i}}e^{\text{i}\varphi_{i}}\ket{E_{i}},
\end{equation}
where $0\le\varphi_{i}\le2\pi$,    $0\leq r_i\leq 1$, $\sum_{i=1}^{3}r_{i}=1$, and $\{\ket{E_i}\}$ stands for a set of Hamiltonian eigenstates such that $\hat{H}\ket{E_{i}}=E_{i}\ket{E_{i}}$. 
The energy spacing defines the transition frequencies according to $\omega_{ij}=-\omega_{ji}=(E_i-E_j)/\hbar$.
Direct calculation gives  
\begin{subequations}\label{means}
\begin{align}
\mean{\op{H}}&=E_{j}+\hbar\sum^3_{i=1}r_i\omega_{ij},\label{Emean} \\
\sigma_{\op{H}}&=\sqrt{\frac{\hbar^2}{ 2}\sum_{i,k=1}^{3}r_ir_k\omega^2_{ik}}.\label{sigma}
\end{align}
\end{subequations}

If $r_j=0$ for some $j$, the vector (\ref{Psi0}) becomes an effective two-level state. 
Similarly, in presence of energy degeneracy such that $E_j=E_k$ for $j\neq k$, $\ket{\psi_0}$  reduces to
\begin{equation}\label{Psi0degA}
\ket{\psi_0}=\sqrt{r_{i}}e^{\text{i}\varphi_{i}}\ket{E_{i}}+\sqrt{r_j+r_k}\ket{E_{jk}},
\end{equation}
with $\ket{E_{jk}}$ the normalized state 
\beq\label{deg23A}
\ket{E_{jk}}=\sqrt{\tfrac{r_j}{r_j+r_k}}e^{\text{i}\varphi_{j}}\ket{E_{j}}+\sqrt{\tfrac{r_k}{r_j+r_k}}e^{\text{i}\varphi_{k}}\ket{E_{k}}
\eeq
with corresponding energy eigenvalue $E_j=E_k$.
Thus, if degeneracy is present among the states $\{\ket{E_i}\}$ in Eq. (\ref{Psi0}) we end up again with an effective qubit state. 
In such case, Eqs. \eqref{means} lead to: 
$\mathcal{E}=\hbar\omega(1-r_i)$, $\mathcal{E}^*=\hbar\omega r_i$ and $\sigma_{\op{H}}^2=\hbar^2\omega^2 r_i(1-r_i)$, %
with $r_i\neq 0,1$ and $\omega$ the single relevant transition frequency. 
This verifies that qubits saturate the Bhatia-Davis inequality. 
Further, for an equally weighted qubit state ($ r_i=\frac{1}{2}$) we have $\alpha=\beta= 1$, and the three bounds coincide (star symbol in Fig.~\ref{fig:alpha-beta}).  

Based on the previous discussion, and with the purpose of analyzing systems with three (non-degenerate) accessible states we will assume, unless otherwise stated, that all three $\{r_i\}$ are nonzero, and that no degeneracies exist so that $E_1<E_2<E_3$.
In this case, Eqs. (\ref{means}) give (after elimination of $r_1$ due to the normalization condition) 
\begin{align}\label{Esig}
\mathcal{E}&=\mean{\op{H}}-E_{1}=\hbar\omega_{21}[r_2+r_3(1+\Omega)],\nonumber\\
\mathcal {E}^{*}&=E_3-\mean{\op{H}}=\hbar\omega_{21}[(1-r_3)(1+\Omega)-r_2],\\
\sigma_{\op{H}}&=\hbar\omega_{21}\sqrt{[r_2+r_3(1+\Omega)^2]-[r_2+r_3(1+\Omega)]^2
}\nonumber,
\end{align}
where we introduced the ratio
\beq\label{Omega}
\Omega\equiv\frac{\omega_{32}}{\omega_{21}}
\eeq
that measures the relative energy separation level.
Substitution of Eqs. (\ref{Esig}) into Eqs. (\ref{cocientes}) gives
\begin{subequations}\label{abc}
\begin{eqnarray}
\alpha&=&\sqrt{\frac{r_{2}+r_{3}\left(1+\Omega\right)^{2}}{\big[r_{2}+r_{3}(1+\Omega)\big]^{2}}-1},\\\label{alpha}
\beta&=&
\frac{\sqrt{r_2+r_3(1+\Omega)^2-[r_2+r_3(1+\Omega)]^2
}}{(1-r_3)(1+\Omega)-r_2}.
\label{beta}
\end{eqnarray}
\end{subequations}
We observe that when expressed in terms of the parameters $\{r_i\}$ and $\Omega$, the Bhatia-Davis inequality is trivially satisfied, since $0\le\alpha\beta\le1$ is equivalent to $0\le\frac{1}{1+\Omega}\le1$. 
%

\section{
The QSL map of a qutrit system}\label{mapaQSL}

\begin{figure}[h]
\includegraphics[scale=.43]{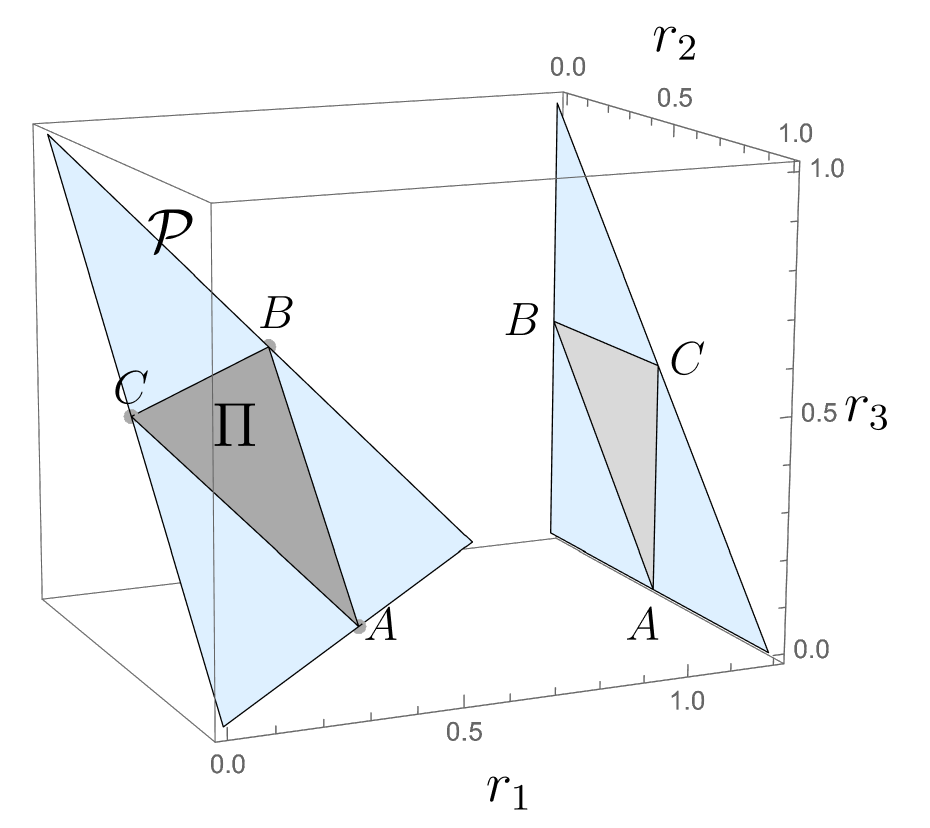}
\caption{All (normalized) initial states $\ket{\psi_0}$ given by Eq. \eqref{Psi0} are mapped to the region $\mathcal P$, conformed by the points $(r_1,r_2,r_3)$ that define a probability distribution. %
The projections on the plane $(r_2,r_3)$ are also shown.}
\label{fig:Plane2Triangle}
\end{figure}

It is clear from Eqs. (\ref{cotas}) and (\ref{abc}) that the bound that determines the QSL depends on both the Hamiltonian (via the relative energy spacing $\Omega$), and the initial state (via the probability distribution $\{r_i\}$).
The set of all triads $(r_1,r_2,r_3)$ that conform a probability distribution lie on the plane $r_1+r_2+r_3=1$, denoted as $\mathcal P$ in Fig. \ref{fig:Plane2Triangle}.
Each initial state of the form (\ref{Psi0}) is thus in correspondence with a point on this plane. 
The gray shadowed region delimited by the vertices $A=(\tfrac{1}{2},\tfrac{1}{2},0)$, $B=(\tfrac{1}{2},0,\tfrac{1}{2})$, and $C=(0,\tfrac{1}{2},\tfrac{1}{2})$, will be of particular interest below, and is denoted with $\Pi$.  
%
The projections of $\mathcal{P}$ and $\Pi$ on the plane $(r_2,r_3)$ are also displayed for future purposes. 

In what follows we elucidate the 
QSL of the states associated to each point in $\mathcal{P}$, given a class of Hamiltonians determined by $\Omega$.   

\subsection{Edges of $\Pi$}

We start considering the points along the edges of $\Pi$, parametrized according to:
 \begin{align}
 \overline{AB}&:=\Bigl\{\Bigl(\tfrac{1}{2},\tfrac{1}{2}-r,r\Bigr)\, \Big\vert\, 0<r<\tfrac{1}{2}\Bigr\},\nonumber\\ 
 \overline{BC}&:=\Bigl\{\Bigl(\tfrac{1}{2}-r,r,\tfrac{1}{2}\Bigr)\, \Big\vert\, 0<r<\tfrac{1}{2}\Bigr\},\\
 \overline{CA}&:=\Bigl\{\Bigl(r,\tfrac{1}{2},\tfrac{1}{2}-r\Bigr)\, \Big\vert\, 0<r<\tfrac{1}{2}\Bigr\}\nonumber.
 \end{align}   
\paragraph{} Points along the edge $\overline{AB}$ give 
\begin{subequations}
\begin{equation}
\alpha=\sqrt{1+\frac{r(1-2r)\Omega^2}{\big(\tfrac{1}{2}+r\Omega\big)^{2}}}>1,
\end{equation}
and
\begin{equation}
\beta=\frac{\sqrt{(1+\Omega)^2-\Omega (1-2r)[2+\Omega(1-2r)]}}{\Omega (1-2r)+(1+\Omega)}<1.
\end{equation}
\end{subequations}
Therefore $\beta<1<\alpha$ for all $\Omega$, and from Eqs. (\ref{cotas}) it follows that 
\beq\label{triad1}
\tau_{\textrm{qsl}}=\tau_{\textrm{ML}}=\frac{\pi}{\omega_{21}(1+2r\Omega)}\quad(\textrm{edge }\overline{AB}).
\eeq

\begin{figure}[h!]
\centering
\includegraphics[scale=0.4]{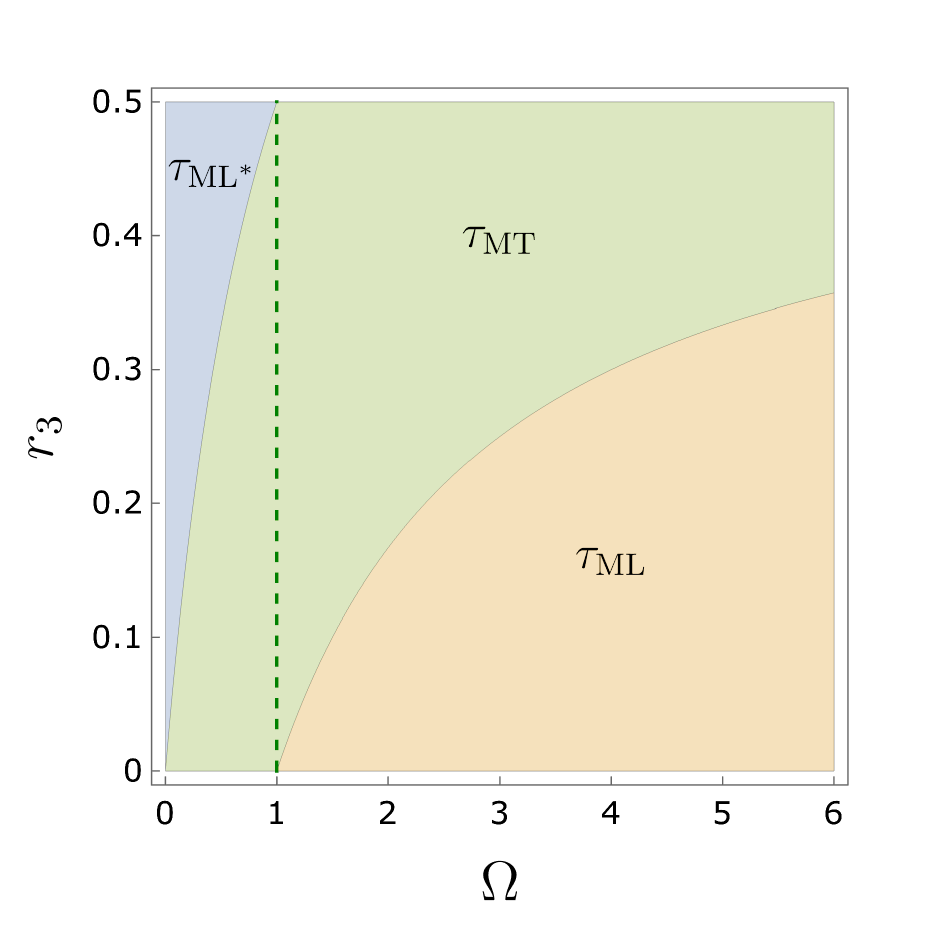}
\caption{Regions in the space $(\Omega,r_3=\frac1{2}-r)$ characterized by the different  bounds that determine $\tau_{\textrm{qsl}}$ along points lying on the edge $\overline{CA}$. For $\Omega=1$ the MT bound dominates along the entire segment.}
\label{reglimites}
\end{figure}

\paragraph{} Points along the edge $\overline{BC}$ correspond to 
\begin{subequations}
\begin{equation}
\alpha=\sqrt{1-\frac{4r(1+2r+2\Omega)}{\big(1+2r+\Omega\big)^{2}}}<1,
\end{equation}
and
\begin{equation}
\beta = \sqrt{1+\frac{4r(1-2r)}{(1-2r+\Omega)^2 }} >1,
\end{equation}
\end{subequations}
hence $\alpha<1<\beta$ for all $\Omega$, and Eqs. (\ref{cotas}) imply that 
\beq\label{triad3}
\tau_{\textrm{qsl}}=\tau_{\textrm{ML*}}=\frac{\pi}{\omega_{21}(1+\Omega-2r)}\quad(\textrm{edge }\overline{BC}).
\eeq
\paragraph{} Points along the edge $\overline{CA}$ correspond to
\begin{subequations}
\begin{equation}\label{alphaCA}
\alpha=\sqrt{1+\frac{(1-2r)(1+\Omega)\big[r(1+\Omega)-1\big]}{\big[1-r+\Omega\bigl(\frac{1}{2}-r\bigr)\big]^{2}}},
\end{equation}
and
\begin{equation}\label{betaCA}
\beta=\frac{\sqrt{-4 r^2 (1+\Omega )^2+4 r (1+\Omega)+\Omega ^2}}{2 r (1+\Omega )+\Omega}.
\end{equation}
\end{subequations}
As argued in the Appendix \ref{Appendix1}, the quantum speed limit along the edge $\overline{CA}$ is thus given by:
\begin{widetext}
\beq\label{qslCA}
\tau_{\textrm{qsl}}=\begin{cases}
			\tau_{\textrm{ML*}}=\dfrac{\pi}{\omega_{21}[\Omega+2r(1+\Omega)]} &\textrm{if} \quad 0<r<\frac{1-\Omega}{2(1+\Omega)}\quad (0< \Omega<1),\\
            \tau_{\textrm{MT}}=\dfrac{\pi}{\omega_{21}\sqrt{\Omega^2+4r(1+\Omega)-4r^2(1+\Omega)^2}} & \textrm{if} \quad \max\Big\{0,\frac{1-\Omega}{2(1+\Omega)}\Big\}<r<\min\Big\{\frac{1}{1+\Omega},\frac{1}{2}\Big\},\\
            \tau_{\textrm{ML}}=\dfrac{\pi}{\omega_{21}[1+(1-2r)(1+\Omega)]} & \textrm{if} \quad \frac{1}{1+\Omega}<r<\frac{1}{2} \quad (1< \Omega).
		 \end{cases}
\eeq
\end{widetext}
Figure \ref{reglimites} shows the three different regions in the space $(\Omega,r_3=\frac1{2}-r)$, according to the bound that gives the QSL in line with Eq. (\ref{qslCA}). 

We observe that for $\Omega<1$ 
only the ML* and the MT bounds
become relevant.
In the equally-spaced case, with $\Omega=1$, the quantum speed limit is always given by $\tau_{\textrm{MT}}$, whereas for $\Omega>1$
either the MT or the ML bound dominates. 
For $\Omega\rightarrow0$ the QSL is given by $\tau_{\textrm{ML*}}$, while it is given by  $\tau_{\textrm{ML}}$ for $\Omega\rightarrow\infty$.

\subsection{Region $\mathcal P$}
\begin{figure}
\centering
\includegraphics[width=0.9\columnwidth]{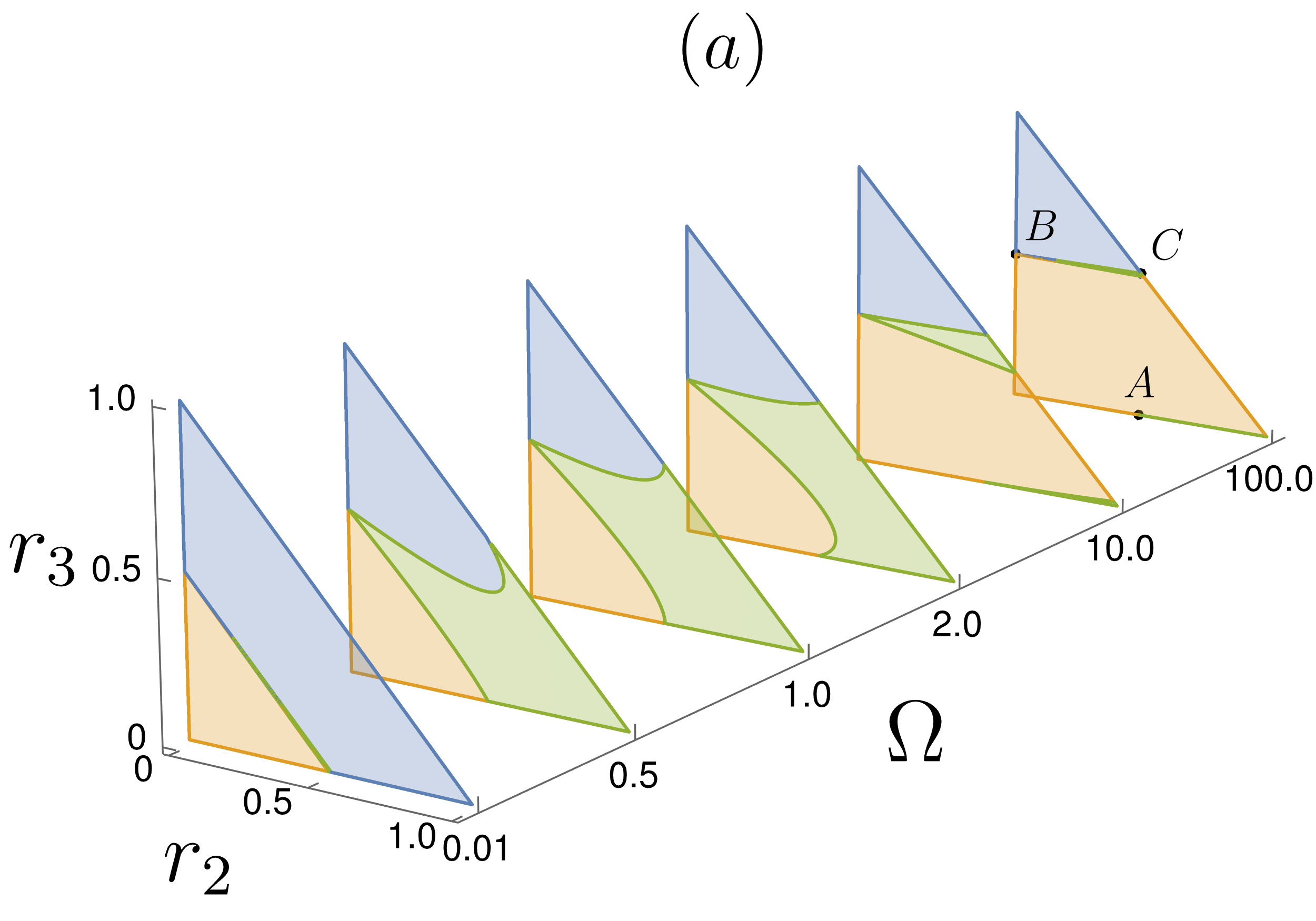}
\includegraphics[width=0.9\columnwidth]{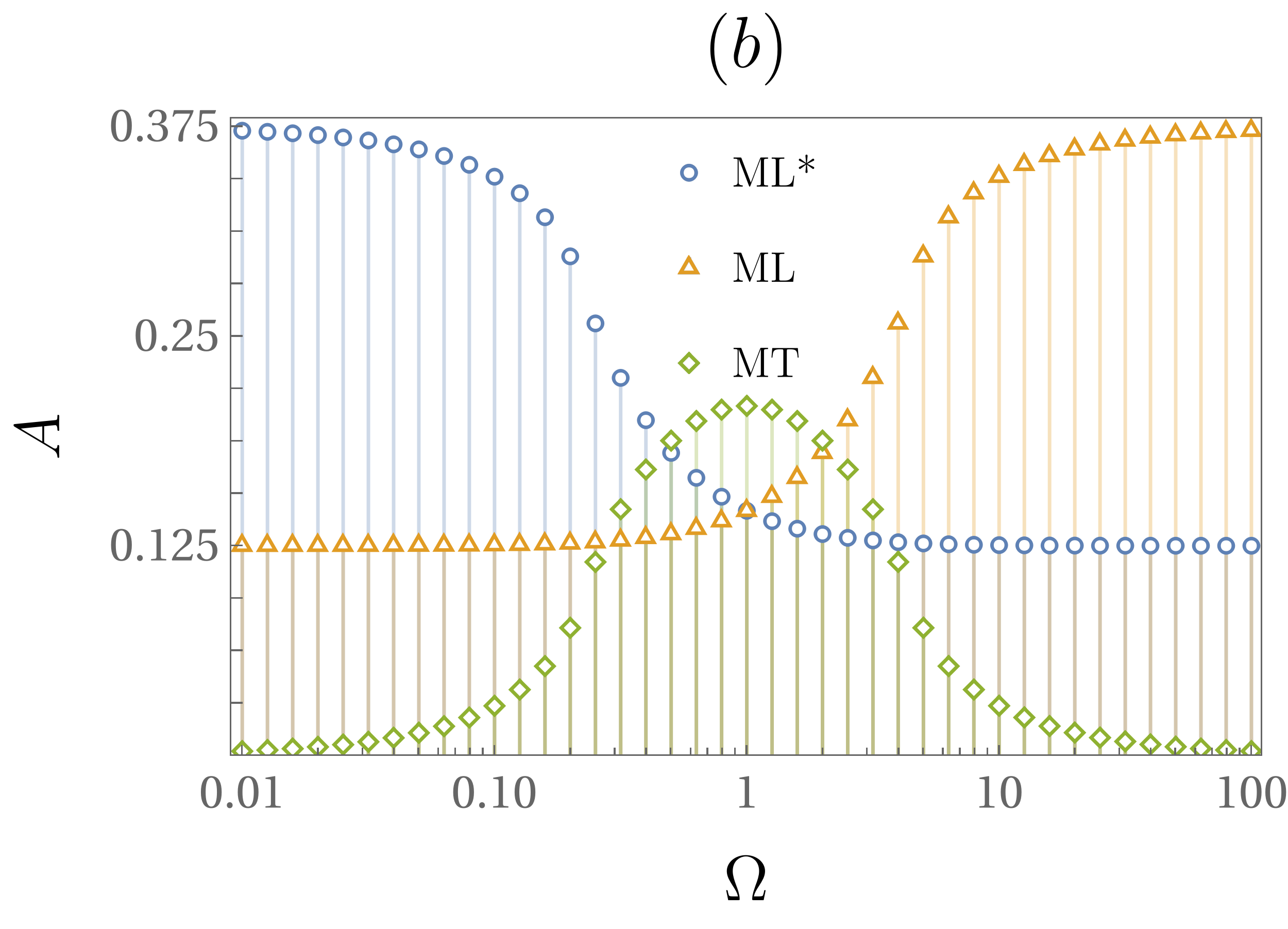}
\caption{Panel (a): Regions in the plane $(r_2,r_3)$, at different values of $\Omega$, colored according to the bound that determines $\tau_{\textrm{qsl}}$ (blue: $\tau_{\textrm{qsl}}=\tau_{\textrm{ML*}}$, orange: $\tau_{\textrm{qsl}}=\tau_{\textrm{ML}}$, and green: $\tau_{\textrm{qsl}}=\tau_{\textrm{MT}}$). Panel (b): Area of each of the three regions of panel (a)
as a function of $\Omega$ (semi-logarithmic scale).}
\label{regeneral}
\end{figure}

As for the rest of points conforming the set $\mathcal P$, we resort to Eqs.~\eqref{cotas} and \eqref{abc} to determine the regions in the space $(r_2,r_3)$ in which $\tau_{\text{qsl}}$ is determined either by $\tau_{\text{ML}}$ (orange), $\tau_{\text{MT}}$ (green), or $\tau_{\text{ML*}}$ (blue). 
Such regions are connected sets and are  continuously transformed as a function of $\Omega$, as shown in Figure \ref{regeneral}(a).  

At the extreme values $\Omega\rightarrow0$ and $\Omega\rightarrow\infty$, $\mathcal{P}$ is divided into two simply connected regions separated by $\overline{AB}$ in the first case, and by $\overline{BC}$ in the second one, and the QSL is given either by the ML or the ML* bound.
These limiting situations correspond to an energy degeneracy:
For 
$\Omega=(\omega_{32}/\omega_{21})\rightarrow0$ the system tends to the degenerate case $E_2=E_3$
, whereas for $\Omega\rightarrow \infty$ the effective degeneracy is $E_1=E_2$. In these cases, as discussed above,
the state becomes an effective two level system when expressed in the basis of non-degenerate energy eigenstates. 
These states saturate the Bhatia-Davis inequality, thus lie along the dashed line in Fig. \ref{fig:alpha-beta}.

%


%
For intermediate values of $\Omega$, the (green) area where the QSL is given by the MT is non-negligible. 
As $\Omega\rightarrow1$, this region and the orange one (ML) grow at the expenses of the blue  (ML*) area. 
At $\Omega=1$ the MT region reaches its maximum extension, approximately given by 0.20833, and starts to diminish monotonically for $\Omega>1$, while the ML area keeps growing until it reaches 3/4 of the total size of $\mathcal{P}$.
Figure \ref{regeneral}(b) shows the area of the three regions as a function of $\Omega$ (in logarithmic scale), thus verifying that indeed the equidistant-energy-level case ($\Omega=1$) gives the largest area for MT. Notice that areas for ML* and ML vary symmetrically with respect to the value $\Omega=1$.

\section{\label{sect:Orthogonality} Attaining orthogonality in the qutrit system}
An initial state \eqref{Psi0} reaches an orthogonal one in a finite time $\tau$, only if certain conditions among the transitions frequencies $\omega_{ij}$ and the probability distribution $\{r_i\}$ are met. 
The analysis of such general and sufficient conditions has been advanced in Ref. \cite{SevillaQRep2021} and 
are revisited here for completeness, but now resorting to elementary geometrical reasonings. 
In addition, knowing the orthogonality times is useful for many information tasks, so the determination of such times in terms of the relevant parameters is desirable.

\begin{figure}[H]
\centering
\includegraphics[scale=0.35]{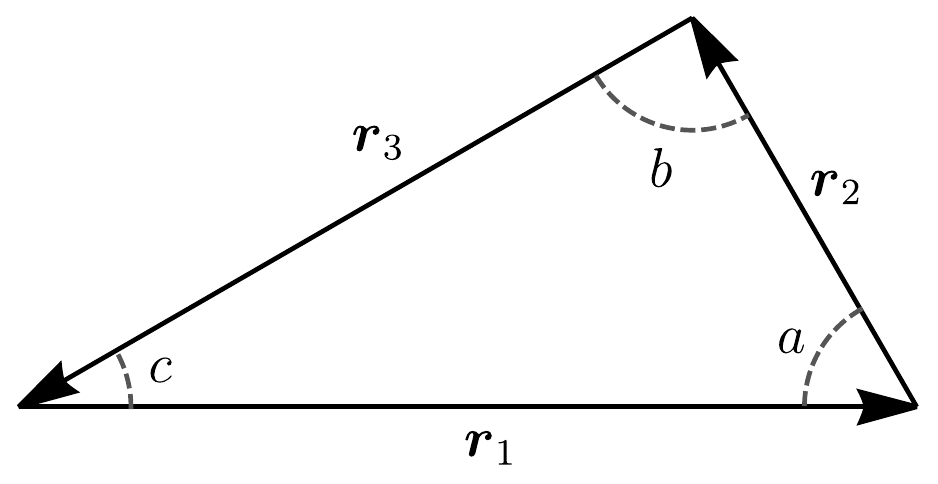}
\caption{Triangle formed by the three vectors $\boldsymbol{r}_i(\tau)$ satisfying the condition (\ref{vector}), indicating that an orthogonal state has been reached.}
\label{triangulo}
\end{figure}

After a time $t$, the initial state $\ket{\psi_0}$ evolves into
\begin{equation}\label{estado}
\ket{\psi(t)}=e^{-\text{i}\hat Ht/\hbar}\ket{\psi_0}=\sum_{i=1}^{3}\sqrt{r_{i}}e^{\text{i}\varphi_{i}}e^{-\text{i}E_{i}t/\hbar}\ket{E_{i}},
\end{equation} 
and attains an orthogonal (perfectly distinguishable) state at time $t=\tau$ provided 
\begin{equation}\label{OverlapGral}
\braket{\psi(\tau)}{\psi(0)}=\sum^3_{i=1}r_i e^{\text{i}E_{i}\tau/\hbar}=0.
\end{equation}
This orthogonality condition is equivalent to the vector equation 
\beq \label{vector}
\sum^{3}_{i=1}\boldsymbol{r}_i(\tau)=\boldsymbol{0},
\eeq
where $\boldsymbol{r}_i$ is a vector in $\mathbb R_2$ defined as $\boldsymbol{r}_i(t)=r_i(\cos \omega_{i1}t,\sin 
\omega_{i1}t)$.
Equation (\ref{vector}) states that an orthogonal state is reached at time $\tau$ provided the three vectors $\boldsymbol{r}_i(\tau)$ form the edges of a triangle of sides $r_1$, $r_2$ and $r_3$, as depicted in Fig. \ref{triangulo}. 

From this simple geometric representation it follows that for the vectors $\boldsymbol{r}_i(\tau)$ to form a triangle, it is necessary that the lengths $r_1,r_2,r_3$ comply with the triangle inequality, $r_i\leq r_j+r_k$. 
This constriction, together with the normalization condition $\sum_ir_i=1$,  
imply that for the triangle to be actually formed, it must necessarily hold that $r_i\leq r_j+r_k=1-r_i$, so
\beq \label{maxri}
r_i\leq \frac{1}{2}
\eeq
for all $i$.
Points lying outside the region $\Pi$ in Fig. \ref{fig:Plane2Triangle} do not comply with this condition, whence we conclude that any initial state that is mapped outside the $\Pi$ region will not attain an orthogonal state, irrespective of the Hamiltonian.

A second conclusion that ensues from purely geometric considerations is that initial states that are mapped into points inside the region $\Pi$ [satisfying equation (\ref{maxri})] can reach an orthogonal state only for specific Hamiltonians. 
Indeed, once such initial state is determined, the length of the triangle sides is fixed. 
Thus, it is readily seen from Fig. (\ref{triangulo}) that the orthogonality condition can only hold for some time $\tau$, provided that the angles $\theta_{ij}(\tau)$ between pairs of vectors $\boldsymbol{r}_i(\tau),\boldsymbol{r}_j(\tau)$ add up to $\pi$.
The fact that the angles $\theta_{ij}(\tau)$ [in Figure \ref{triangulo}: $\theta_{12}(\tau)=a$, $\theta_{23}(\tau)=b$, $\theta_{31}(\tau)=c$] depend on the transition frequencies $\omega_{ij}$ (the relative angular velocity at which each vector rotates), makes evident that only for certain values of such frequencies, i.e., only for specific Hamiltonians, the orthogonality condition can be satisfied. 
This observation will be rigorously stated below. 

The orthogonality condition (\ref{OverlapGral}) allows for many (in fact infinite) solutions for $\tau$, so the system evolves between orthogonal states recurrently.
We will refer to the time it takes the state to transform into an orthogonal state for the first time as the {\it first orthogonality time}, defined as $\tau_{1}=\min\{\tau\,|\,\braket{\psi(\tau)}{\psi(0)}=0\}$. 

By inspection of Fig. \ref{triangulo}, the internal angles of the triangle are given in terms of $\tau_1$ and the transition frequencies as follows   
\begin{eqnarray}\label{internalangles}
a&=&\pi-\omega_{21}\tau_1,\nonumber\\
b&=&\pi-\omega_{32}\tau_1,\\
c&=&\pi-a-b=\omega_{31}\tau_1-\pi.\nonumber
\end{eqnarray}
The well-known relations between the internal angles of a triangle and its sides relate $\omega_{21}$, $\omega_{32}$ and  $\tau_1$ with the coefficients $\{r_i\}$ according to:
\begin{eqnarray}\label{internalangles2}
\omega_{21}\tau_1&=&\pi-\arccos\biggl({\frac{r_1^2+r_2^2-r_3^2}{2r_1r_2}}\biggr),\nonumber\\
\omega_{32}\tau_1&=&\pi-\arccos\biggl({\frac{r_2^2+r_3^2-r_1^2}{2r_2r_3}}\biggr),\\
\omega_{21}\tau_1
(1+\Omega)&=&\pi+\arccos\biggl({\frac{r_1^2+r_3^2-r_2^2}{2r_1r_3}}\biggr).\nonumber
\end{eqnarray}

\subsection{Orthogonality times}
 
As we have just seen, whether an initial state represented by a point in the $\Pi$ region in Fig. \ref{fig:Plane2Triangle} attains or not an orthogonal state will depend on the specific Hamiltonian via the transition frequencies, $\omega_{21}$, $\omega_{32}$, since these must satisfy Eqs.~\eqref{internalangles} consistently with Eqs.~(\ref{internalangles2}) (see also Eq.~(\ref{sines}) below). 
Now, when the necessary conditions for evolving towards an orthogonal state are met, a fundamental question is to determine the precise time at which this occurs. In what follows we briefly review the resulting orthogonality times in different regions of the $\Pi$ region, previously reported in \cite{SevillaQRep2021}.

In \cite{SevillaQRep2021} it was verified that the vertices $A,B$ and $C$, corresponding each to an equally weighted  qubit state, reach an orthogonal state for all Hamiltonians (i.e., irrespective of $\{\omega_{ij}\}$). 
These states are also the fastest, complying with  $\tau_1=\tau_\textrm{qsl}=\tau_\textrm{ML}=\tau_\textrm{ML*}=\tau_\textrm{MT}$ \cite{LevitinPRL2009}. The orthogonality times of states represented by these vertices are 
\beq\label{tauABC}
\tau_{\textrm{vert}}=\frac{(2l+1)\pi}{\omega},
\eeq
with $l=0,1,\dots$, and $\omega$ the frequency connecting the two energy eigenstates contributing to $\ket{\psi}$. The last equation indicates that the system visits the orthogonal state at odd multiples of $\tau_1$.

In was also demonstrated in \cite{SevillaQRep2021} that points along the edges $\overline{AB}$, $\overline{BC}$ and $\overline{CA}$ correspond to states that reach an orthogonal one provided the transition frequencies are commensurable.
Specifically, orthogonality is reached whenever 
\beq\label{Omeganm}
\Omega=\frac{\omega_{32}}{\omega_{21}}=\frac{n}{m},
\eeq
with $n,m$ positive integers such that: 
$n=2(l'+1)$ and $m=2l+1$ for the segment $\overline{AB}$;
$n=2l'+1$ and $m=2(l+1)$ for the segment $\overline{BC}$; 
$n=2l'+1$ and $m=2l+1$ for the segment $\overline{CA}$, with $l,l'=0,1,\dots$.
The orthogonality times in all cases are 
\beq\label{tausedges}
\tau_{\textrm{edges}}=\frac{m\pi}{\omega_{21}}.
\eeq

For the points inside the region $\Pi$ in Fig. \ref{fig:Plane2Triangle}, the orthogonality time, the probabilities $\{r_i\}$ and the energy-level separations are related in a more involved way. 
Their relation is determined by the law of sines applied to the triangle in Fig. \ref{triangulo}, namely
\beq \label{sines}
\frac{r_1}{\sin \omega_{32}\tau}=\frac{r_2}{-\sin \omega_{31}\tau}=\frac{r_3}{\sin \omega_{21}\tau},
\eeq
which gives \cite{SevillaQRep2021}
\begin{equation}\label{sol}
r_{i}=\frac{\sin\omega_{jk}\tau}{\sin\omega_{31}\tau+
\sin\omega_{12}\tau+\sin\omega_{23}\tau}, 
\end{equation}
where the indices $(i,j,k)$ are taken in a cyclic permutation of $(1,2,3)$.
In this case, orthogonality is reached for a very particular Hamiltonian (or set $\{\omega_{ij}\}$) once the initial state (the point $(r_1,r_2,r_3)$) is fixed.
Similarly, for a given Hamiltonian, a point $(r_1,r_2,r_3)\in\Pi$ can be found such that an orthogonal state is attained at some $\tau$.

Equation (\ref{sol}) is, however, not useful for determining the first orthogonality time, $\tau_1$.
The latter can be obtained by combining Eqs. (\ref{internalangles}) and (\ref{internalangles2}), which give, after elimination of $r_1$:
\begin{eqnarray}\label{taurelleno}
\tau_1&=&\tfrac{1}{\omega_{21}}\Big\{\pi - \arccos\Big[ \tfrac{1-2r_3}{2r_2(1-r_2-r_3)}-1 \Big]\Big\},\nonumber \\
\tau_1&=&\tfrac{1}{\Omega\omega_{21}}\Big\{\pi - \arccos \Big[ \tfrac{2(r_2+r_3)-1}{2r_2r_3} -1 \Big]\Big\}, \\
\tau_1&=&\tfrac{1}{(1+\Omega)\omega_{21}}\Big\{\pi + \arccos \Big[ \tfrac{1-2r_2}{2r_3(1-r_2-r_3)} -1 \Big]\Big\}.\nonumber 
\end{eqnarray}
For a given Hamiltonian (fixed $\omega_{21}$ and $\Omega$), the above expressions establish a system of equations [from the third line in (\ref{internalangles}) it follows that only two of the equations in (\ref{taurelleno}) are independent] whose solutions $r_2,r_3$ determine a curve on the projected region $\Pi$, representing the initial states that evolve towards a distinguishable one at $t=\tau_1$.
Below we show such curves for different values of the transition frequencies.

\section{Map of speeds}
\label{sect:MapSpeeds}

Having identified the quantum speed limit 
and the first orthogonality time in the different regions of the parameter space 
, we now proceed to analyze the speed of evolution of qutrit systems that evolve into distinguishable states.
With that purpose we focus on the ratio
\beq\label{Speed}
s\equiv\frac{\tau_{\textrm{qsl}}}{\tau_1}.
\eeq
This figure of merit  quantifies the (dimensionless) speed of evolution by comparison of the speed at which a first orthogonal state is reached (as measured by $\tau^{-1}_{1}$), relative to the maximum speed  permitted by the quantum speed limit (as measured by $\tau^{-1}_{\text{qsl}}$).
Since $\tau_{1}$ is bounded according to $0<\tau_{\text{qsl}}\leq \tau_{1}$, $s$ lies in the interval $(0,1]$.
In line with the definition of $s$, a system state is said to evolve faster (or to have a higher speed of evolution) if the time it takes to firstly reach an orthogonal state is closer to the minimal possible time 
$\tau_{\text{qsl}}$
, determined by the system's own energetic resources. 
The fastest evolution, for which  orthogonality is attained in the least possible time, corresponds to $\tau_{1}=\tau_{\text{qsl}}$ and $s=1$, whereas the slowest evolution occurs when the initial state does not transform into an orthogonal one at all, so $\tau_{1}\rightarrow\infty$, resulting in $s\rightarrow 0$. 

As stated above
, for initial states mapped into the vertices $A,B$ and $C$ it holds that $s=1$.  
The speed of any other point in the region $\Pi$ (containing the points that can reach an orthogonal state in finite time) will  be analyzed in what follows.

\subsection{Speed along the edges}
\label{sect:EdgeSpeeds}
\begin{figure*}
\centering
\includegraphics[width=0.64\columnwidth]{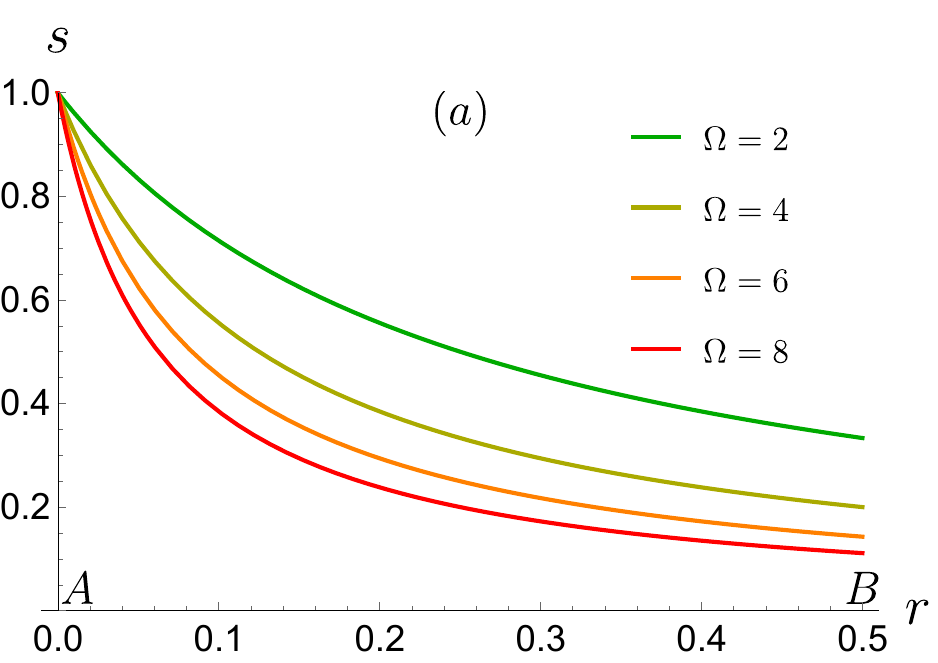}
\includegraphics[width=0.64\columnwidth]{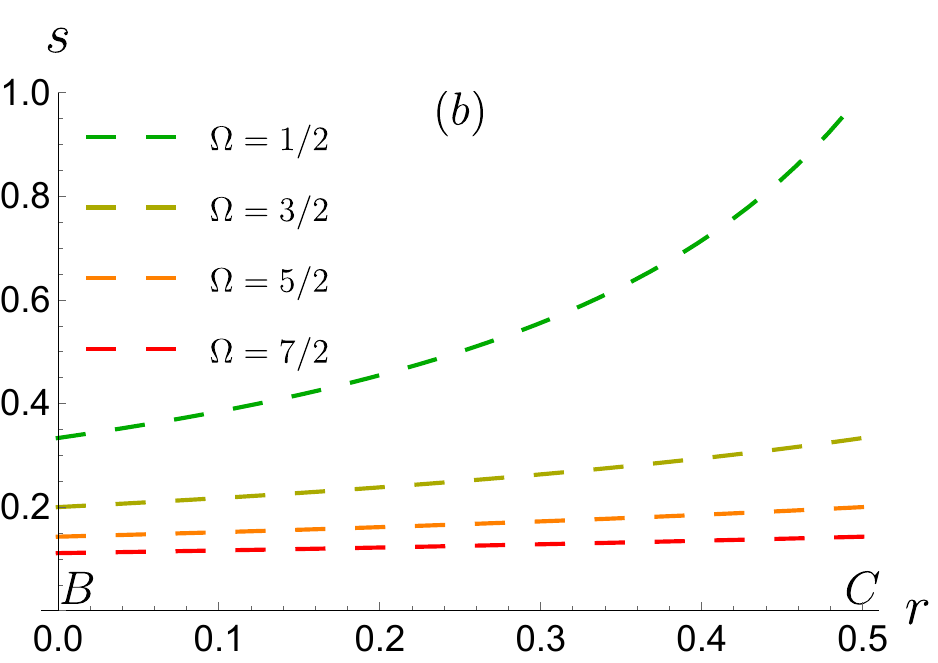}
\includegraphics[width=0.64\columnwidth]{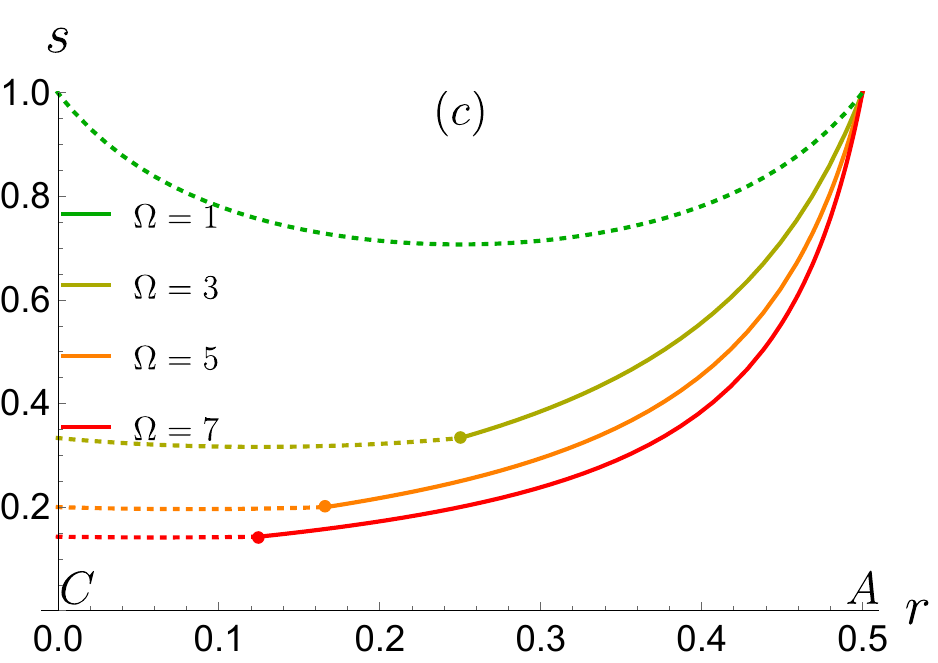}
\caption{Speed towards orthogonality of points (states) lying along the edges $\overline{AB}$ (panel a), $\overline{BC}$ (panel b), and $\overline{CA}$ (panel c). Solid lines correspond to $\tau_{\textrm{qsl}}=\tau_{\textrm{ML}}$, dashed lines to $\tau_{\textrm{qsl}}=\tau_{\textrm{ML*}}$, and dotted lines to $\tau_{\textrm{qsl}}=\tau_{\textrm{MT}}$.}
\label{smap}
\end{figure*}

\paragraph{} For points along the edge $\overline{AB}$, we resort to Eq. (\ref{tausedges}) to determine $\tau_1$, and to Eq. (\ref{triad1}) giving the QSL along this edge, and obtain
\begin{equation}\label{sab}
s=\frac1{1+2r\Omega},\quad\Omega=2,4,6,...,
\end{equation}
where the discreteness of $\Omega$ ensues from the restrictions on $n$ and $m$ that guarantee the orthogonality, discussed below Eq. (\ref{Omeganm}).
The speed (\ref{sab}) is a decreasing function of $r\in(0,\frac{1}{2})$, and is bounded as
\beq\label{boundsab}
\frac{1}{1+\Omega}<s<1.
\eeq
This implies that for initial states mapped into the segment $\overline{AB}$, the fastest configuration corresponds to Hamiltonians such that $\Omega=2$, with speeds $s\in(\tfrac{1}{3},1)$ [see Fig. \ref{smap}(a)].

\paragraph{} Analogously, for points along the edge $\overline{BC}$, Eqs. (\ref{tausedges}) and (\ref{triad3}) result in
\begin{equation}
s=\frac1{2(1+\Omega-2r)}\quad \Omega=\frac{1}{2},\frac{3}{2},\frac{5}{2},\dots,
\end{equation}
where again the restriction on $\Omega$ ensues from Eq. (\ref{Omeganm}) and the conditions established just below it.
Now $s$ is an increasing function of $r$, bounded as 
\beq\label{boundsbc}
\frac{1}{2(1+\Omega)}<s<\frac{1}{2\Omega}.
\eeq
The optimal $\Omega$, the one that gives the highest speed along the segment $\overline{BC}$, is thus $\Omega=1/2$. This gives $\tfrac{1}{3}<s<1$, as in the previous case [see Fig.~\ref{smap}(b)].

\paragraph{} For points along the edge $\overline{CA}$, 
Eq.~\eqref{tausedges} gives $\tau_1=\pi/\omega_{21}$, provided $\Omega=1,3,5,\dots$.
%
From here and resorting to Eq. (\ref{qslCA}) we see that the quantum speed limit is either given by $\tau_{\textrm{MT}}$ or $\tau_{\textrm{ML}}$, but never by $\tau_{\textrm{ML*}}$.
The relative evolution speed is thus given by
\begin{widetext}
\beq\label{sCA}
s=\begin{cases}
            \dfrac{1}{\sqrt{\Omega^2+4r(1+\Omega)-4r^2(1+\Omega)^2}} & \textrm{if}\quad 0<r<\frac{1}{1+\Omega}\quad ( \Omega=1,3,5,\dots),\\
            \dfrac{1}{[1+(1-2r)(1+\Omega)]} &\textrm{if}\quad \frac{1}{1+\Omega}<r<\frac{1}{2} \quad (\Omega=3,5,\dots).
		 \end{cases}
\eeq
\end{widetext}
In this case $s$
reaches its maximum value ($s=1$) as $r$ approaches $\frac{1}{2}$ regardless of $\Omega$, and attains the minimal value of $s = \frac{1}{\sqrt{1+\Omega^2}}$ at $r = \frac{1}{2(1+\Omega)}$.
Therefore, the fastest configuration along this segment correspond to $\Omega = 1$, giving $\frac{1}{\sqrt{2}} < s < 1$ [see Fig.~\ref{smap}(c)].

In Fig.~\ref{smap} the speed $s$ for each of the edges is shown as $r$ is varied in the interval $(0,\frac{1}{2})$, for  different values of $\Omega$.
Solid, dashed and dotted lines indicate that the corresponding QSL is given by $\tau_{\textrm{ML}}$, $\tau_{\textrm{ML*}}$, and $\tau_{\textrm{MT}}$, respectively.
Recall that for all vertices it holds $s=1$, so the curves in Fig.  \ref{smap} have a discontinuity at the points indicating the vertices $B$ and $C$.


\subsection{Speed within the region $\Pi$}\label{regionpi}

In order to compute $s=\tau_{\text{qsl}}/\tau_1$ for points inside the central triangle $\Pi$ in Fig. \ref{fig:Plane2Triangle}, 
we need to identify $\tau_{\text{qsl}}$ with either one of the bounds ML, MT, or ML*, for each point in $\Pi$.
To this aim, we make the following
\begin{theorem*}
The quantum speed limit for states with $r_i<1/2$ that do reach orthogonality is always given by $\tau_{\textrm{MT}}$.
\end{theorem*}
To prove this, we resort to Eq. (\ref{sol}) to write the solutions $r_i$ in terms of $\Omega$ and the dimensionless first orthogonality time $\tilde\tau_1\equiv\omega_{21}\tau_1$, obtaining
\begin{subequations}\label{solbis}
\begin{eqnarray}
r_{2}&=&\frac{\sin[(1+\Omega)\tilde\tau_1]}{\sin[(1+\Omega)\tilde\tau_1]-
\sin\tilde\tau_1-\sin\Omega\tilde\tau_1},\\
r_{3}&=&\frac{-\sin\tilde\tau_1}{\sin[(1+\Omega)\tilde\tau_1]-
\sin\tilde\tau_1-\sin\Omega\tilde\tau_1},
\end{eqnarray}
\end{subequations}
and $r_1=1-r_2-r_3$. 

Further, imposing the condition that all $a,b$ and $c$ in Eq. (\ref{internalangles}) lie between $0$ and $\pi$ gives the following equations:
$0<\tilde\tau_1<\pi$, $0<\tilde\tau_1<\frac{\pi}{\Omega}$, and $\frac{\pi}{1+\Omega}<\tilde\tau_1<\frac{2\pi}{1+\Omega}$.
This leads directly to
\beq\label{xrestricted}
\frac{\pi}{1+\Omega}<\tilde\tau_1<\min\Big\{\pi,\frac{\pi}{\Omega},\frac{2\pi}{1+\Omega}\Big\}=\begin{cases}
			\pi, & \Omega<1,\\
            \frac{\pi}{\Omega}, & \Omega>1.
		 \end{cases}
\eeq
In the allowed interval for $\tilde\tau_1$ it can be verified that $\alpha$ and $\beta$ in Eqs. (\ref{abc}), with $r_2,r_3$ given by (\ref{solbis}), are both less than unity, whence $\tau_{\textrm{qsl}}=\tau_{\textrm{MT}}$. \qedsymbol

This means, in particular, that if a point in $\Pi$ reaches orthogonality, then it must be found in the (green) region dominated by the MT bound in Fig. \ref{regeneral}(a). 

Now that we have identified $\tau_{\textrm{qsl}}=\tau_{\textrm{MT}}$ for points in $\Pi$ attaining an orthogonal state, we can determine the speed of evolution (\ref{Speed}).
The quantum speed limit $\tau_\text{MT}=\pi\hbar/2\sigma_{\hat H}$ is directly computed from Eq. (\ref{sigma}) with $\{r_i\}$ determined by (\ref{taurelleno}); then $s$ is obtained resorting to any of the (equivalent) expressions for 
$\tau_1$ in (\ref{taurelleno}). 
From the numerical solution of equations (\ref{taurelleno}) we find that:
\begin{theorem*}
For a given $\Omega$, the set of solutions $\{r_i\}$ of Eqs.~\eqref{taurelleno} draws a curve within the region $\Pi$ that goes from a vicinity of vertex $B$ towards a vicinity of vertex $A$ whenever $\Omega<1$, and towards vertex $C$ whenever $\Omega>1$. For $\Omega=1$ the curve ends at $(\frac{1}{4},\frac{1}{2},\frac{1}{4})$, the middle point of the edge $\overline{CA}$.
\end{theorem*}

\begin{figure}[H]
\centering
\includegraphics[width=0.9
\columnwidth, trim=15 15 10 10]{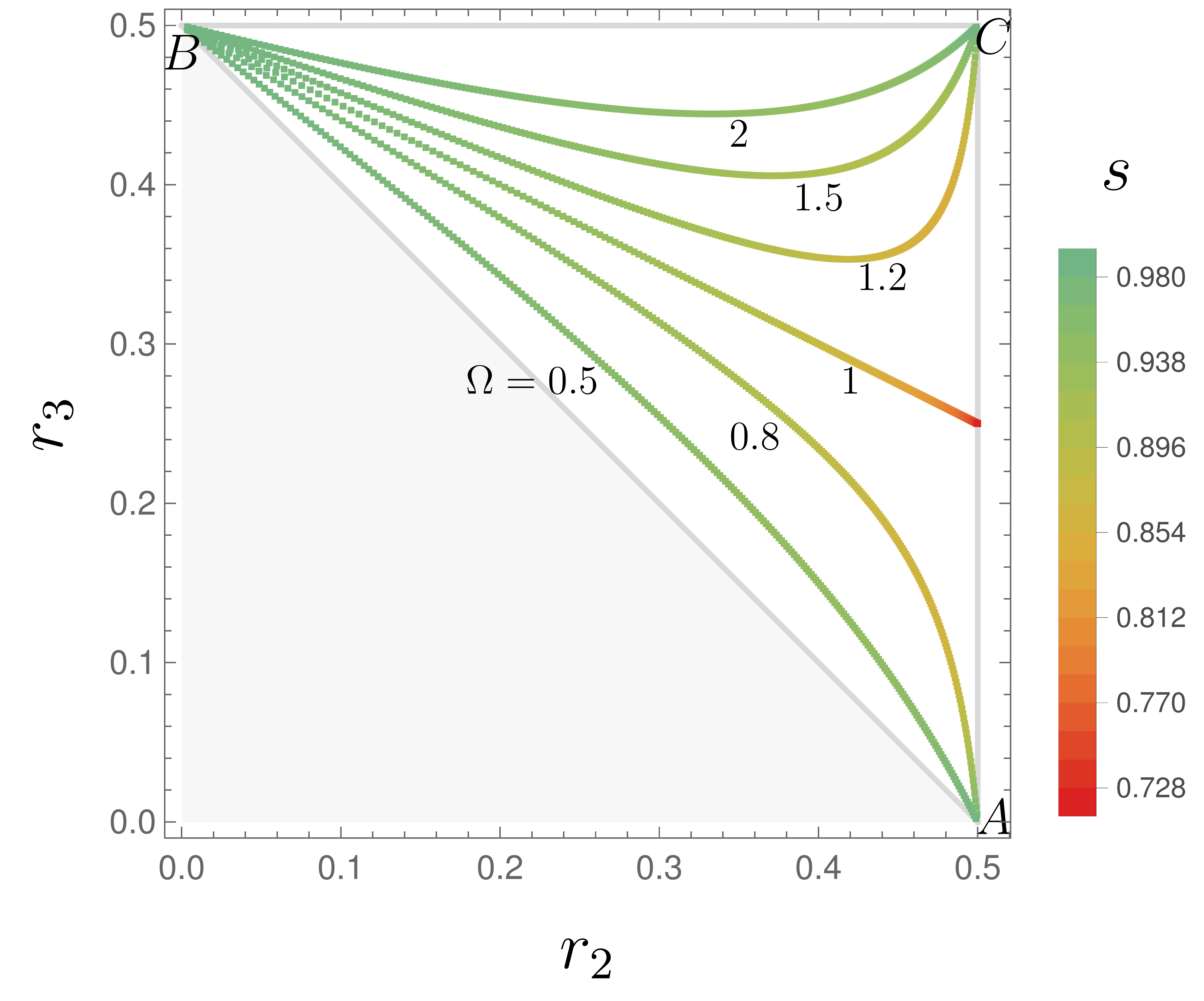}
\caption{For fixed $\Omega$ (here we considered the values 0.5, 0.8, 1, 1.2, 1.5 and 2) the points in the space $(r_2,r_3)$ corresponding to distributions $\{r_i\}$ that reach an orthogonal state in a finite time are displayed. The color scale indicates the associated speed $s$, given by Eq.  \eqref{Speed}.}
\label{relleno ortogonalidad}
\end{figure}

Figure \ref{relleno ortogonalidad} depicts the resulting speed $s$ (color scale) along the curves consistent with the orthogonality condition [points $(r_2,r_3)$ that solve Eq. (\ref{taurelleno})], for different values of $\Omega$.  

We observe that in line with the above assertion,  
for arbitrary $\Omega<1$ there exists a family of initial states that transform into an orthogonal one in a finite time, and continuously connect the qubit state $\ket{q_{B}}=\tfrac{1}{\sqrt 2}(\ket{E_1}+e^{\text{i}\varphi_3}\ket{E_3})$ (corresponding to the vertex $B$) with the qubit $\ket{q_{A}}=\tfrac{1}{\sqrt 2}(\ket{E_1}+e^{\text{i}\varphi_2}\ket{E_2})$ (corresponding to the vertex $A$). 
Analogously, for arbitrary $\Omega>1$ there exists a family of initial states that evolve into an orthogonal one, continuously connecting the qubit states $\ket{q_{B}}$ and $\ket{q_{C}}=\tfrac{1}{\sqrt 2}(\ket{E_2}+e^{\text{i}\varphi_3}\ket{E_3})$ (corresponding to the vertex $C$).

Interestingly, the speeds in Fig. \ref{relleno ortogonalidad} are relatively high in general (no less than $\approx$ 72\% of the maximum speed allowed by the QSL).

\section{Analysis of a concrete physical system}\label{physical}
The results obtained in the previous sections are now applied to a system of two interacting bosons that can hop between two sites \cite{BarriosPhysicaA2022}. 
The dynamics under consideration is governed by the Hamiltonian
\begin{multline}
\op{H}=-J(\op{a}_0^{\dagger}\op{a}_1+\op{a}_1^{\dagger}\op{a}_0)-K(\op{a}_0^{\dagger}\op{a}_0^{\dagger}\op{a}_1\op{a}_1+\op{a}_1^{\dagger}\op{a}_1^{\dagger}\op{a}_0\op{a}_0)\\
+U\bigl[\op{n}_0(\op{n}_0-1)+\op{n}_1(\op{n}_1-1)\bigr].
\end{multline}
Here $J>0$ and $K>0$ denotes the energy due to the transport of one and two bosons, respectively, that transit between the two sites, and $U>0$ stands for the intensity of the repulsive on-site energy.
Further, $\op{a}^\dagger_\mu$ and $\op{a}_\mu$ are the creation and annihilation operators of bosons in site $\mu=0,1$, and $\op{n}_\mu=\op{a}^\dagger_\mu\op{a}_\mu$ is the number operator that gives the number of bosons in site $\mu$.

The system's Hilbert space has dimension 3, and is spanned by the Fock basis
\begin{eqnarray}\label{Fock}
\ket{n}&\equiv& \ket{2-n,n}=\ket{2-n}_0\otimes \ket{n}_1\nonumber\\
&=&\frac{\bigl(\op{a}_{0}^{\dagger}\bigr)^{2-n}}{\sqrt{(2-n)!}}\frac{\bigl(\op{a}_{1}^{\dagger}\bigr)^{n}}{\sqrt{n!}}\ket{\text{vac}},
\end{eqnarray} 
where $\ket{n'}_{\mu}$ is the state with $n'$ particles in the site $\mu$ (and is eigenstate of the number operator $\op{n}_\mu$),  and  $\ket{\text{vac}}$ is the two-mode vacuum state. 
The basis (\ref{Fock}) reads explicitly
\beq\label{Fockbasis}
\bigl\{\ket{n}\bigr\}=\bigl\{\ket{0}=\ket{2,0},\ket{1}=\ket{1,1},\ket{2}=\ket{0,2}\bigr\},
\eeq
and corresponds to states in which the number of bosons is conserved (so while there are  $n'=0,1,2$ bosons in one site, the other one contains the remaining $2-n'$ particles).

\begin{figure}[H]
\centering
\includegraphics[scale=0.5]{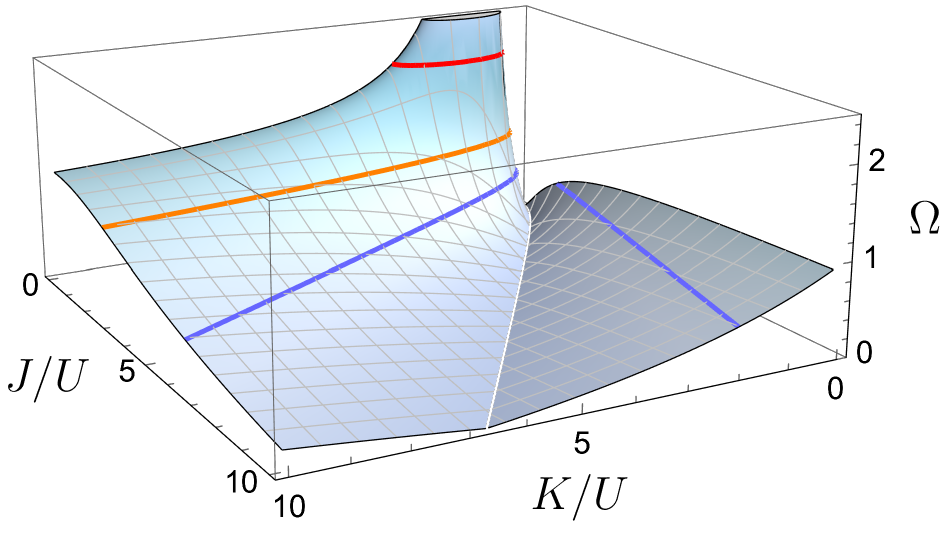}
\caption{Relative energy separation level $\Omega$ [Eq. \eqref{Omega}] as a function of $J/U$ and $K/U$ for the (generalized) Bose-Hubbard Hamiltonian (\ref{matrixH}). The blue level curve corresponds to $\Omega=1/2$, the orange one to $\Omega=1$, and the red one to $\Omega=2$. These correspond to the values of $\Omega$ that maximize the speed $s$ along the edges $\overline{BC}$, $\overline{CA}$, and $\overline{AB}$ (see Fig. \ref{smap}).}
\label{omegaBH}
\end{figure}

The Hamiltonian matrix in the basis \eqref{Fockbasis} reads
\begin{equation}\label{matrixH}
H =
\begin{pmatrix}
2U & -\sqrt{2}J & -2K\\
-\sqrt{2}J & 0 & -\sqrt{2}J\\
-2K & -\sqrt{2}J & 2U
\end{pmatrix}.
\end{equation}
After direct diagonalization the following (ordered) eigenvalues are obtained:
\begin{subequations}
\begin{eqnarray}
E_1&=&U-K-\sqrt{4J^2+(K-U)^2},\\
E_2&=&\begin{cases}
			2(U-K)-E_1 & \textrm{if}\,\;J^2<2K(U+K),\\
            2(U+K) &\textrm{if}\,\; J^2>2K(U+K),
		 \end{cases}
   \\
E_3&=&\begin{cases}
			2(U+K) &\textrm{if}\,\; J^2<2K(U+K),\\
            2(U-K)-E_1 &\textrm{if}\,\; J^2>2K(U+K).
		 \end{cases}
\end{eqnarray}
\end{subequations}
From here $\Omega$ can be written explicitly in terms of $J$, $K$ and $U$ as
\beq\label{Omega-BH}
\Omega
=\begin{cases}
			\frac{U+3K-\sqrt{4J^2+(K-U)^2}}{2\sqrt{4J^2+(K-U)^2}} &\textrm{if}\,\; J^2<2K(U+K),\\
            \frac{-(U+3K)+\sqrt{4J^2+(K-U)^2}}{U+3K+\sqrt{4J^2+(K-U)^2}} &\textrm{if}\,\; J^2>2K(U+K).
		 \end{cases}
\eeq
The condition $J^2=2K(U+K)$ indicates a crossing of energy levels (for which $\Omega=0$), and divides the surface $\Omega(J/U,K/U)$ into two regions as shown in Fig. \ref{omegaBH}, separated by the white line (indicating the condition of vanishing $\Omega$).
Since, as seen in section \ref{Aalphabeta}, this corresponds to an effective qubit state, we will pay no attention to this case and assume that $J^2\neq2K(U+K)$.

The level curves $\Omega(J/U,K/U)=\Omega_0$ (with $\Omega_0$ constant) determine the values of $J/U$ and $K/U$ that, together with $r_2$ and $r_3$, give rise to maps of QSL according to Fig.~\ref{regeneral}(a). 
As has been pointed out there, $\Omega=1$ separates two distinct physical situations, 
one for which the ML* or the MT bound specifies the QSL for the majority of states in $\mathcal{P}$ ($0<\Omega<1$), and the other one ($\Omega>1$) in which the the role of ML and ML* are symmetrically exchanged. 
For the particular physical system considered here, the manifold $\Omega(J/U,K/U)$ defined by Eq.~\eqref{Omega-BH} splits into a manifold for which $0<\Omega<1$ for $J^2>2K(U+K)$ (rightmost region in Fig.~\ref{omegaBH}), and into another one for which $\Omega$ is unbounded, when $J^2<2K(U+K)$ (leftmost region in Fig. \ref{omegaBH}).

As shown in the previous section, the states that reach an orthogonal one in a finite time are contained either in the set $\Pi$, each having a QSL given by the MT bound, or on its edges for specific values of $\Omega$, with different QSL according to the analysis of Sect.~\ref{sect:EdgeSpeeds}.
Thus, for the rightmost region of Fig.~\ref{omegaBH} orthogonality can only be reached for points inside the region $\Pi$ (area contained between the edge $\overline{AB}$ and the line $\Omega=1$ in Fig.~\ref{relleno ortogonalidad}), or along the edge $\overline{BC}$ provided $\Omega=1/2$ (with QSL given by $\tau_{\textrm{ML*}}$, as seen in Fig. \ref{smap}(b)).
The blue level curve in Fig. \ref{omegaBH} indicates the case with $\Omega=1/2$. 
Notice that the standard Bose-Hubbard Hamiltonian (corresponding to $K=0$ and thus trivially located at the rightmost region) leads to $\Omega=1/2$ whenever $J=\sqrt{2}U$. 
Thus, when this condition is met and the initial state has $r_3=1/2$ (so the associate point lies in $\overline{BC}$), the evolution towards an orthogonal state is guaranteed, with the maximal speed $s=\tau_{\textrm{ML*}}/\tau_1$ corresponding to the green dashed line in Fig. \ref{smap}(b).   

For the leftmost region of Fig. \ref{omegaBH}, $\Omega$ is unbounded, thus in this manifold the whole set $\Pi$ in Fig.~\ref{relleno ortogonalidad} is available for points $(r_2,r_3)$ to reach orthogonality,
as well as points along the edges $\overline{BC}$, $\overline{CA}$ and $\overline{AB}$,  provided $\Omega$ is semi-integer, odd and even, respectively. 
The corresponding (blue, orange, and red) level curves are shown in Fig. \ref{omegaBH}.

\begin{figure}[H]
\centering
\includegraphics[scale=0.45]{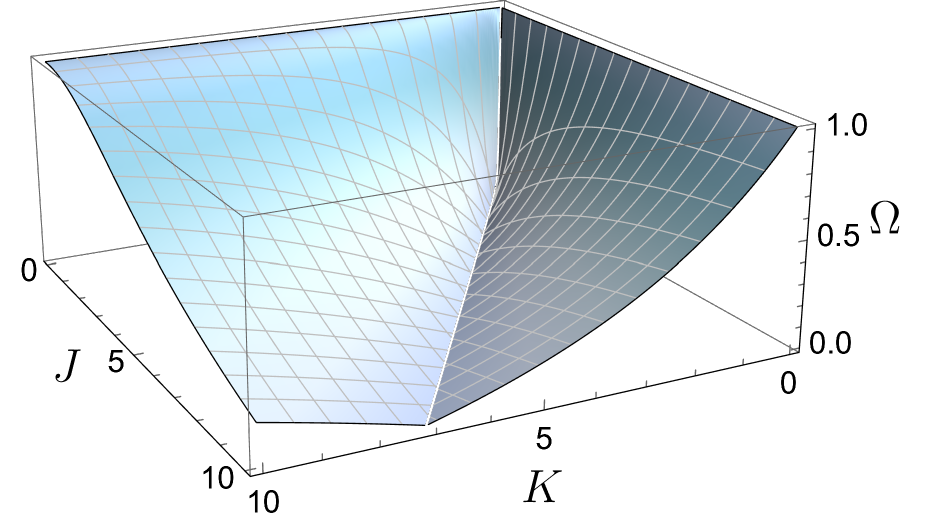}
\caption{Relative energy separation level $\Omega$ as a function of $J$ and $K$ for the Hamiltonian (\ref{matrixH}) with $U=0$. }
\label{BHUcero}
\end{figure}

For $U=0$ the relative energy separation level $\Omega(J,K)$ is depicted in Fig. \ref{BHUcero}.
Again two regions appear, depending on whether $J^2>2K^2$ (rightmost surface) or $J^2<2K^2$ (leftmost surface), divided by the line $J^2=2K^2$, for which $\Omega=0$.
In contrast with the case $U\neq 0$, now  $0\leq \Omega\leq 1$ for all $J,K$. Therefore the quantum speed limit for $U=0$ is either $\tau_{\textrm{MT}}$, for points within $\Pi$, or $\tau_{\textrm{ML*}}$, for points along $\overline{BC}$ provided $\Omega=1/2$. 

For $U=0$ the matrix (\ref{matrixH}) takes the form
\begin{equation}\label{matrixHbis}
H_0 =\begin{pmatrix}
0 & h_1 & h_2\\
h_1 & 0 & h_1\\
h_2 & h_1 & 0
\end{pmatrix},
\end{equation}
with $h_1=-\sqrt{2}J$, and $h_2=-2K$. 
The Hamiltonian matrix (\ref{matrixHbis}) has an structure that serves also to describe other physical systems, such as a single particle that can hope among the wells of a 3-well potential. 
The energetic cost of the tunneling is $h_1$ if the transition connects adjacent wells ($1\leftrightarrow 2$, or $2\leftrightarrow 3$), and $h_2$ if the transition involves the farthest apart sites ($1\leftrightarrow 3$).
If the three wells form a ring array and all transitions between first neighbours are equivalent, then $h_1=h_2$, meaning $J=\sqrt{2}K$, and the resulting $\Omega$ vanishes (degenerate case).  
These last observations show that the previous analysis may be of use in a wide range of physical scenarios.


\section{Final remarks}

We have presented a comprehensive analysis of the quantum speed limit in a closed system of three levels that evolves unitarily under an arbitrary time-independent Hamiltonian. 

For a fixed ratio of the energy-level separations $\Omega$, we found the explicit partition of the set of energy-probability distributions $\{r_i\}$ into the subsets $\{r_i\}_\textrm{MT}$, $\{r_i\}_\textrm{ML}$ and $\{r_i\}_\textrm{ML*}$, whose elements define pure states with quantum speed limit given by the Mandelstam-Tamm, the Margolus-Levitin and by the recently established dual bound, respectively. Such partition induces the color map, depicted in Fig. \ref{regeneral}(a), which graphically identifies each of the bounds' domain, 
%
determined by the preparation of the initial states. This 
finds application, for example, in systems of ultracold atoms, where many parameters can be adjusted to control the initial
state as well as the dynamics of the system \cite{GrossScience2017}. 

The role of the probability of the lowest (highest)-energy state in determining the QSL of $\ket{\psi_0}$ was elucidated for an arbitrary Hamiltonian.
When the lowest (highest) energy state is the most probable one, i.e.,  $r_1\geq1/2$ ($r_3\geq1/2$), the ML (ML dual) bound governs the system's evolution speed. 
However, the identification of the quantum speed limit is highly non-trivial
in the region for which $r_1<1/2$ (equivalently $r_2+r_3>1/2$) and $r_3<1/2$, since the QSL depends now on $\Omega$, as shown in Fig. \ref{regeneral}(a).
%

We proved that only for $r_{1,2,3}<1/2$ (states mapped into the interior of the central triangle $\Pi$) orthogonality can be reached in a finite time for an appropriate set of transition frequencies, in which case $\tau_{\textrm{qsl}}$ is given by the Mandelstam-Tamm bound.
These results provide a criterion to discern whether an initial configuration may result in a transit between distinguishable states and, together with the area distribution in Fig. \ref{regeneral}(b), indicate that the number of such configurations increases as the energy levels tend to be equally-spaced.  
The observation that the Mandelstam-Tamm bound is the dominant one in qutrit systems 
that reach an orthogonal state, endows $\tau_\textrm{MT}$ with a prominent role over the ML and its dual bounds.

For configurations that attain an orthogonal state, we computed a detailed map of the evolution speed, identifying continuously connected families of initial states represented by the curves in Fig. \ref{relleno ortogonalidad}, whose orthogonality time is no less than 72$\%$ of the QSL. 
The speed of evolution was also elucidated for the edges of $\Pi$ (see Fig. \ref{smap}). In this case we found that the speed increases as the relative energy separation level $\Omega$ decreases. 

\begin{figure}[H]
\centering
\includegraphics[scale=0.3]{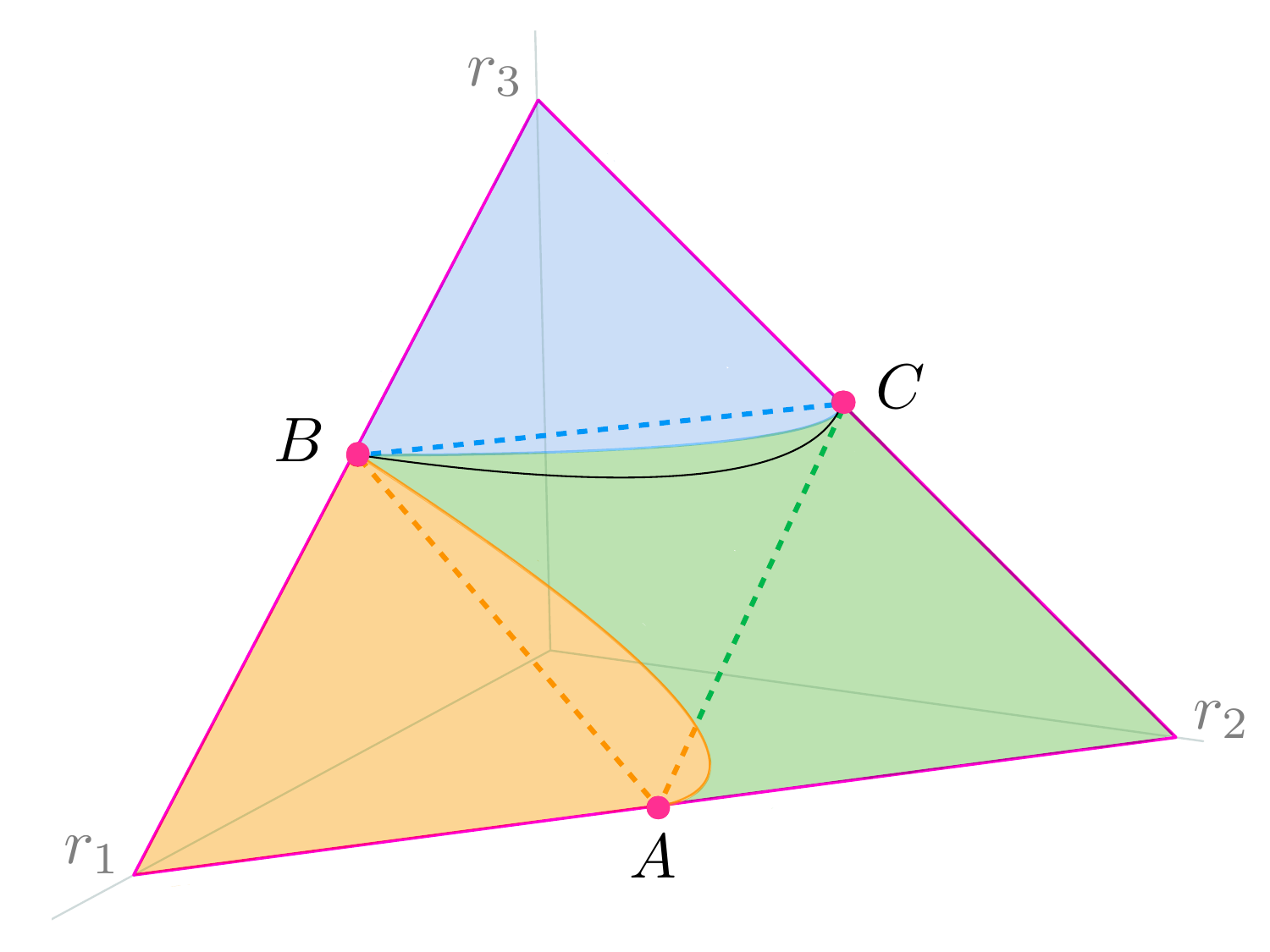}
\caption{Qualitative graphical summary of the quantum speed limit characterization (MT: green, ML: orange and ML$^*$: blue) in the parameter space, and the identification of regions indicating the initial states that may evolve towards an orthogonal state 
(see text for details).}
\label{summary}
\end{figure}

A  qualitative graphical summary of the resulting characterization in the space $(r_1,r_2,r_3)$ is depicted in Fig. \ref{summary}. 
The solid magenta lines enclosing the big triangle correspond to effective qubit states. Magenta dots (vertices $A,B$ and $C$) point out the only qubit states that reach an  orthogonal state, irrespective of the Hamiltonian and with the maximal evolution speed ($s=1$). 
Dotted lines, connecting the vertices $A, B$ and $C$, distinguish states that transform into an orthogonal one only for certain commensurable transition frequencies [see Eq. (\ref{Omeganm})].
Points lying outside the inner triangle $ABC$ (that defines the region $\Pi$) correspond to initial states that do not reach an orthogonal one irrespective of the Hamiltonian, while points inside it may attain orthogonality only for specific values of $\Omega=\omega_{32}/\omega_{21}$.
The color indicates 
the quantum speed limit, 
$\tau_{\text{qsl}}=\tau_{\text{MT}}$ in the green region, $\tau_{\text{qsl}}=\tau_{\text{ML}}$ in the orange region, and $\tau_{\text{qsl}}=\tau_{\text{ML*}}$ in the blue one. 
The color map  
changes as the Hamiltonian varies (i.e., is $\Omega$ dependent) only inside the parallelogram delimited by the vertices $A,B,C$ and the point $(0,1,0)$. The QSL map 
inside this region
in Fig. \ref{summary} schematizes the case $\Omega=2$ [see the fourth slice in Fig. \ref{regeneral}(a)], and 
the solid black line represents the set of initial states that evolve towards a distinguishable one for that value of $\Omega$ [see the upper curve in Fig. \ref{relleno ortogonalidad} for the precise form of the curve and its corresponding speed].

%
Extensions of the problem considered here to larger systems is of clear interest, since high-dimensional systems provide a larger space to store and process information. 
An extension of the present geometrical approach in this direction is thus desirable. 
In brief, the generalization to $n$ dimensions of the region $\mathcal{P}$ in Fig.~\ref{fig:Plane2Triangle}  corresponds to the $(n-1)$-dimensional simplex $\mathcal{P}_n$ [the set of $n$-tuples $(r_1,\ldots,r_n)$ that satisfy $0\le r_i\le1$ and $\sum_{i=1}^n r_i=1$]. 
We expect that the corresponding generalization of the $\Pi$ region would correspond to an embedded $(n-1)$-dimensional manifold containing the $n$-tuples associated with the states that reach orthogonality in a finite time, given appropriate Hamiltonian parameters, strongly restricted. 
Regarding the determination of these latter states, the vector equation (\ref{vector}) (directly generalizable to the $n$-dimensional case), implies that the $n$ vectors $\boldsymbol{r}_{i}(\tau)$ form a $n$-sided polygon (irregular in general).
This imposes constrains on the values of the $n-1$ frequencies $\omega_{i1}$, and consequently on the Hamiltonian for which orthogonality can be attained. 
For those frequencies that satisfy such constriction, we expect a manifold of dimension $n-2$ that contains states that eventually evolve into an orthogonal one, for a fixed Hamiltonian.   
Whether the Mandelstam-Tamm bound continues to dominate over such manifold is a conjecture that deserves future exploration. Indeed, the complete identification of the regions where $\tau_\textrm{qsl}$ is determined either by $\tau_\textrm{MT}$,  $\tau_\textrm{ML}$ or $\tau_{\textrm{ML}^*}$, is foreseeably highly nontrivial, and will be presented elsewhere. 

Our theoretical analysis is general enough and does not depend on the peculiarities of the physical system, thus offering guidance for determining the conditions under which a specific transformation may induce transitions between perfectly distinguishable states.  
We have made explicit predictions for particular three-level systems amenable to experimental verification: 
a system of two interacting bosons in two accessible sites, governed by an extended Bose-Hubbard Hamiltonian, and a single particle capable of hopping between wells in a three-well potential. 
For the former case, we found that the possibility of having two bosons hopping between wells (as opposed to the standard Bose-Hubbard Hamiltonian in which $K=0$), 
broadens the regions in the parameters space in which an evolution towards an orthogonal state is possible, and either one of three the bounds $\tau_{\textrm{MT}}, \tau_{\textrm{ML}}$, and $\tau_{\textrm{ML*}}$ can be identified with the quantum speed limit, for appropriate values of the parameters. 

In summary, we offered a thorough characterization of the quantum speed limit in qutrit systems, contributing to a deeper understanding of the intrinsic time scales governing quantum unitary evolutions, and of the hierarchy and relevance of the MT, ML and dual bounds. 
Our approach paves the way for further investigations considering more complex quantum systems, providing a framework to explore how energetic resources and initial state configurations shape the dynamics of quantum evolution.

\appendix

\section{Quantum speed limit along the edge $\overline{CA}$}\label{Appendix1}
From Eq. (\ref{alphaCA}) it can be seen that 
for $\Omega\le1$ it holds that $\alpha<1$, whereas for $\Omega>1$ we find three cases: 
\begin{subequations}\label{alfa}
\begin{align}
 \alpha>1& \;\;\text{for} \;\; \frac{1}{1+\Omega}<r<\frac{1}{2},\\
 \alpha=1&\;\;\text{for} \;\; r=\frac{1}{1+\Omega},\\
 \alpha<1& \;\;\text{for} \;\; 0<r<\frac{1}{1+\Omega}. 
 \end{align}
\end{subequations} 

Analogously, from Eq. (\ref{betaCA}) we find that
\begin{subequations}\label{beta2}
\begin{align}
 \beta<1& \;\;\text{for} \;\; \frac{1-\Omega}{2(1+\Omega)}<r<\frac{1}{2},\\
 \beta=1&\;\;\text{for} \;\; r=\frac{1-\Omega}{2(1+\Omega)},\\
 \beta>1& \;\;\text{for} \;\; 0<r<\frac{1-\Omega}{2(1+\Omega)}. 
 \end{align}
\end{subequations} 
Combination of these equations with Eqs. (\ref{cotas}) leads to (\ref{qslCA}).

\section*{Author contributions}
J.E.G: Writing, review \& editing, visualization, validation, methodology, investigation, analysis. F.J.S and A.V.H: Writing, review \& editing, visualization, validation, 
methodology, investigation, analysis, conceptualization, supervision, funding acquisition. 
\section*{Data availability}
Data is provided within the manuscript.
\section*{Funding}
This work was supported by projects UNAM-PAPIIT IN112723 and UNAM-PAPIIT IN112623. J.E.G acknowledges SECIHTI through funding for graduate studies.

\bibliographystyle{apsrev}

\end{document}